\documentclass[aps,pra,reprint,showpacs,amssymb,amsmath,longbibliography]{revtex4-1}
\usepackage{graphicx}

\def\be{\begin{equation}}
\def\ee{\end{equation}}
\def\ber{\begin{eqnarray}}
\def\eer{\end{eqnarray}}
\def\bern{\begin{eqnarray*}}
\def\eern{\end{eqnarray*}}

\def\rv{\mathbf{r}}

\def\pv{\mathbf{p}}
\def\qv{\mathbf{q}}

\def\0v{\mathbf{0}}
\def\1v{\mathbf{1}}
\def\2v{\mathbf{2}}
\def\3v{\mathbf{3}}

\def\ii{\mathrm{i}}

\usepackage[usenames]{color}
\usepackage[normalem]{ulem}

\begin{document}

\title{Role of the kinematics of probing electrons in electron energy-loss
spectroscopy of solid surfaces}
\author{V.~U.~Nazarov}
\affiliation{Research Center for Applied Sciences, Academia Sinica, Taipei 11529, Taiwan}
\email{nazarov@gate.sinica.edu.tw}

\author{V. M. Silkin}
\affiliation{Departamento de F\'{i}sica de Materiales, Facultad de Ciencias Qu\'{i}imicas, Universidad del Pais Vasco/Euskal Herriko Unibertsitatea, Apdo. 1072, San Sebasti\'{a}n/Donostia, 20080 Basque Country, Spain}
\affiliation{Donostia International Physics Center (DIPC), Paseo Manuel de Lardizabal 4, San Sebasti\'{a}n/Donostia, 20018 Basque Country, Spain}
\affiliation{IKERBASQUE, Basque Foundation for Science, 48013 Bilbao, Spain}

\author{E. E. Krasovskii}
\affiliation{Departamento de F\'{i}sica de Materiales, Facultad de Ciencias Qu\'{i}imicas, Universidad del Pais Vasco/Euskal Herriko Unibertsitatea, Apdo. 1072, San Sebasti\'{a}n/Donostia, 20080 Basque Country, Spain}
\affiliation{Donostia International Physics Center (DIPC), Paseo Manuel de Lardizabal 4, San Sebasti\'{a}n/Donostia, 20018 Basque Country, Spain}
\affiliation{IKERBASQUE, Basque Foundation for Science, 48013 Bilbao, Spain}

\begin{abstract}
Inelastic scattering of electrons incident on a solid surface 
is determined by the two properties: (i) electronic response of the target system and (ii) the detailed
quantum-mechanical motion of the projectile electron inside and in the vicinity of the target.
We emphasize the equal importance of the second ingredient, pointing out  the fundamental limitations of the 
conventionally used theoretical description
of the electron energy-loss spectroscopy (EELS) in terms of the ``energy-loss functions''.
Our approach encompasses the dipole and impact scattering as specific cases, with the emphasis on the
quantum-mechanical treatment of the probe electron.
Applied to the high-resolution EELS of Ag surface, our theory largely agrees with recent experiments, while some instructive exceptions are rationalized.
\end{abstract}


\maketitle

\section{Introduction}
Electron energy-loss spectroscopy (EELS) is an efficient and widely used experimental method
to study excitation processes on clean and adsorbates-covered surfaces of solids, and in thin (including atomically thin) films \cite{Hillier-44,Ibach-82,Eberlein-08,Egerton-09}. 
This method utilizes the inelastic scattering of electrons,
resulting in both the energy and momentum transfer from the projectiles to diverse kinds of excitations in the  samples. Reflected or transmitted electrons are analyzed with respect to the energy and momentum loss they have experienced in the interaction with a target, revealing a wealth of information about the properties of the latter.

Much efforts have been exerted over years to complement EELS experimental
techniques with comprehensive theoretical pictures \cite{Pines-66,Ritchie-57,Bennet-70,Ibach-82,Tsuei-90-0,Liebsch-98,Liebsch,Nazarov-15}.
In this way, a clear understanding of elementary excitations (such as electron-hole pairs generation, collective electronic excitations -- plasmons, atomic vibrational modes, {\em etc.}), including their momentum dispersion,  for solid surfaces, interfaces, and in thin films have been achieved.

Presently, the main approach to
interpret EELS data theoretically is to use {\em energy-loss functions}. 
A clear example is the surface energy-loss function of a semi-infinite solid,
which, with the neglect of the momentum dispersion, can be written  as \cite{Liebsch}
\begin{equation}
L_s(\omega)= - \text{Im} \frac{1}{\varepsilon(\omega)+1},
\label{Ls}
\end{equation}
where $\varepsilon(\omega)$ is the frequency-dependent dielectric function (DF) of the bulk solid.
This example exhibits an important  feature common also to other, much more sophisticated, loss-functions:
$L_s(\omega)$ of Eq.~(\ref{Ls}) is a property of the target only. Indeed, it is not concerned 
with the setup of the EELS experiment, such as the angles of incidence and reflection (or transmission),
the energy of the electrons in the incident beam, and, which is subtler, the detailed, desirably quantum-mechanical, motion of the probe electrons both outside and inside the target.
As a clear reason why such an approach may not be adequate, we note that it cannot, in principle, 
determine the relative intensities of the surface and the bulk plasmons in a given EELS setup, the bulk response being given in the same approximation by another  energy-loss function
$L_b(\omega)= - \text{Im} \dfrac{1}{\varepsilon(\omega)}$. For systems where the bulk and the surface excitations
overlap, as is the case, e.g., of silver, this constitutes a serious limitation.

Meanwhile, a  theoretical approach to  EELS taking  full account
of the incident electron kinematics had been introduced two decades ago \cite{Nazarov-95S}. 
This is based on the solution to the  problem of the energy-loss by an electron
traveling in the lattice potential of a target, utilizing 
the method  known in the scattering theory as the {\em distorted-wave approximation} \cite{Taylor}
(see Eq.~(\ref{dsdwdW}) of the next section). 
That  formal theory of the  response of the target system coupled 
to the quantum-mechanical motion of the projectile electron has, however, never been 
implemented to the full extent in calculations for specific systems. Indeed, the formalism importantly stipulates
that the density-functional theory\cite{Kohn-65} (DFT) potential used in the calculations
of the ground-state and of the response of the target, on one hand, and the potential which determines the motion
of the projectile electron, on the other,  should be the same crystalline potential. 
Only two specific applications of the theory have been made so far. In the first,
the theory of Ref.~\onlinecite{Nazarov-95S} has been implemented for jellium within a model of the incident electron reflected from an infinite barrier
at a given position above/below the surface \cite{Nazarov-99}.
In the other, which is an application to the inelastic low-energy electron diffraction (LEED) of simple metals, 
a  severe approximation of the kinematic diffraction theory was used \cite{Nazarov-01}.
At the same time,  detailed measurements in the high-resolution EELS (HREELS) of silver surface in the wide energy range have become available recently\cite{Politano-13-3}, 
calling for the implementation of refined theoretical methods.

\begin{figure} [h] 
\includegraphics[width=  \columnwidth, trim= 40 0 40 0, clip=true]{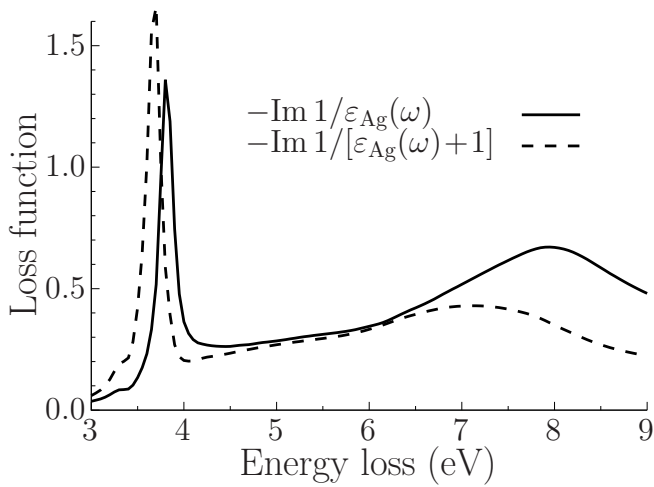}
\caption{\label{sp0} Bulk (solid line) and surface (dashed line) energy-loss functions of silver.
The experimental optical dielectric function  $\varepsilon_{\text{Ag}}(\omega)$ is used \cite{Palik}.
}
\end{figure}

The purpose of this paper is, therefore, two-fold. First, we aim at the implementation of the  theory
of EELS of Ref.~\onlinecite{Nazarov-95S} in its original form, i.e.,
that would treat the incident electron and the electrons of the target system on the same footing.
Secondly, we apply this theory to the EELS of Ag surface, which is exactly the case
when the interplay of the response of the target with the details of the probe's motion 
is especially important, due to the overlap of the bulk and the surface 
features in the excitation spectrum of this material, as is illustrated in Fig.~\ref{sp0}.
Thereby we both further advance the theory of EELS and achieve  an improvement in the understanding of the experimental spectra of the Ag surface.

Since the fully {\em ab initio} solution to the problem of the dielectric response of $d$-metals
still remains a computationally formidable task, we have to resort to some model considerations.
First, we substitute the three-dimensional (3D) problem with a one-dimensional (1D) one,
neglecting the system's non-uniformity in the surface plane. Second, the $d$-electrons 
are included in a phenomenological way, using the model of Liebsch \cite{Liebsch-95,Liebsch-98}
of the background DF. This work should, therefore,
be considered as a step forward toward the full-featured 3D implementation of the same method,
treating also $d$-electrons {\em ab initio}. However, the main ingredients of the theory
(among them, importantly, the necessary inclusion of the optical potential) are presented and discussed
in this work, facilitating  the future implementation of the method in full.

The paper is organized as follows. In Sec.~\ref{form} we remind and further work out  details
of including the motion of the scattered electrons 
in the theory of EELS. In Sec.~\ref{res}, results of the calculations conducted
with the use of our theory are presented and discussed.
Conclusions are collected in Sec.~\ref{concl}.
In the Appendix we detail on some important properties of the model utilized in the calculations.
We use atomic units ($e^2=\hbar=m_e=1$) throughout unless otherwise indicated.

\section{Formalism}
\label{form}

A formal solution to the problem of the inelastic scattering of an electron in the EELS setup,
which incorporates the detailed quantum-mechanical motion of the projectile, 
can be quite generally written as \cite{Nazarov-95S,Nazarov-01}
\begin{widetext}
\begin{equation}
\frac{d^2 \sigma}{d \omega d \Omega} (\pv'\leftarrow \pv) = 
-\frac{16 \pi^3 p'}{p} {\rm Im} \int \frac{\rho_{ext}^*(\rv'')}{|\rv''-\rv|} \chi(\rv,\rv',\omega) \frac{\rho_{ext}(\rv''')}{|\rv'-\rv'''|}  d \rv d \rv' d\rv'' d\rv''',
\label{dsdwdW}
\end{equation}
\end{widetext}
where the left-hand side of Eq.~(\ref{dsdwdW}) is the differential  cross-section
of the scattering from the state of the momentum $\pv$ to a state of the momentum within the solid
angle $d \Omega$ around $\pv'$, and to lose the energy within $d \omega$ around $\omega=(p^2-p'^2)/2$.
In the right-hand side of Eq.~(\ref{dsdwdW}), $\chi(\rv,\rv',\omega)$ is the interacting-electrons 
\footnote{In contrast to the Kohn-Sham $\chi_0$, $\chi$ includes electron-electron interactions,
and the two response functions are 
related as $\chi^{-1}=\chi_0^{-1} - 1/|\rv-\rv'|-f_{xc}$, where $f_{xc}$
is the xc kernel \cite{Gross-85}. }
density-response function of the target,
the complex-valued ``external charge density'' 
\begin{equation}
\rho_{ext}(\rv)=\langle \rv |\pv^+\rangle^*  \, \langle \rv|{\pv'}^-\rangle,
\label{rhor}
\end{equation}
is determined  by the {\em elastic} scattering {\em incoming} and {\em outgoing} wave-functions, 
$|\pv^+\rangle$ and $|\pv^-\rangle$, respectively, which are the solutions to the Lippmann--Schwinger equation \cite{Taylor}
\begin{equation}
\langle \rv |\pv^\pm\rangle=\langle \rv |\pv\rangle+G^0 \! \left(\frac{p^2}{2}\pm \ii 0_+ \right) V_l(\rv) \langle \rv | \pv^\pm\rangle,
\end{equation}
where
$G^0(E)$ is the non-interacting Green's function, $V_l(\rv)$ is the single-particle static
lattice potential,
and $\langle \rv |\pv\rangle=e^{\ii \pv \cdot \rv}/(2\pi)^{3/2}$ are plane-waves. 
\footnote{In Refs.~\onlinecite{Nazarov-95S,Nazarov-01} the solution was given in the momentum space,
of which Eq.~(\ref{dsdwdW}) is the Fourrier transform.}

The structure of Eq.~(\ref{dsdwdW}) has a transparent physical interpretation:
The external charge $\rho_{ext}$ creates an external Coulomb potential,
which, through the density-response function $\chi$, induces the charge fluctuation in the target.
Finally, the Coulomb potential of that fluctuation  couples to the external charge itself, 
causing the inelastic scattering of the latter.

It must be, however, emphasized that the above picture  is no more than a convenient verbal
description of the strict mathematical formalism presented in Ref.~\onlinecite{Nazarov-95S}: The derivation of Eq.~(\ref{dsdwdW}) does not rely on the substitution of the true quantum-mechanical scattering problem for an electron with an artificial charge-density.
It rather solves the problem of the combined elastic and inelastic scattering 
of a charge at an arbitrary many- (or a few) body system,
which, with a mild assumption that the impinging electron  can be considered distinguishable 
from those of the target, can be put into the terms of the density-response function
of the target and the elastic scattering states of the projectile. 
Obtained within the distorted-wave approximation\cite{Taylor}, 
Eq.~(\ref{dsdwdW}) is exact to the first order in the inelastic processes
(the first Born approximation)
and it is exact to {\em all} orders in the elastic scattering.
It includes both  the long- and the short-range
interaction of the probe electron with the target, i.e., the dipole and impact scattering\cite{Liebsch},
respectively, within, most importantly, the quantum-mechanical treatment of the probe itself.

Of course, practically, the quality of specific calculations by Eq.~(\ref{dsdwdW})
depends on the accuracy of the approximations used to calculate its ingredients, i.e., the density-response function of the target $\chi$ and  the wave-functions of the incoming and outgoing electron
utilized  in the construction of $\rho_{ext}$.
We now turn to the use of  specific models.

\subsection{Model of a laterally uniform target}
\label{flat}
In this work we will use a simplification of the potential $V_l(\rv)$=$V_l(z)$ averaged in the plane
parallel to the surface (which
is chosen as the $xy$-plane, with the $z$-axis normal to the surface and directed into vacuum).
In this case the wave-functions are plane-waves in the direction parallel to the surface
\begin{equation}
\langle \rv | \pv^\pm \rangle=   \langle z | p_z^\pm \rangle \frac{ e^{\ii \pv_\|\cdot \rv_\|}}{2\pi} , 
\label{eq5} 
\end{equation}
where the subscript `parallel' denotes the $xy$-projection of a vector.
To take advantage of the scattering theory framework, 
in the following we represent the target with a sufficiently thick slab,
with vacuum both above and below, which is  also consistent with our numerical implementation of the method.
Then $\langle z| p_z^+ \rangle$ can be conveniently found as a solution
to the Schr\"{o}dinger equation with the following asymptotic boundary conditions  
\begin{equation}
\langle z| p_z^+ \rangle=  \frac{1}{2\pi} \left\{
\begin{array}{ll}
 a^+ e^{\ii p_z \, z}+b^+ e^{-\ii p_z \,z} , & z\to +\infty, \\
 c^+ e^{\ii p_z \, z}+d^+ e^{-\ii p_z \,z} , & z\to -\infty,
\end{array}\right. 
\label{eq6}
\end{equation}

\begin{equation}
\begin{split}
& p_z>0 : c^+=1, \ b^+ = 0, \\
& p_z<0 : a^+=1, \ d^+=0.
\end{split}
\label{eq61}
\end{equation}
The asymptotic of $\langle z |p_z'^- \rangle$ is easily obtained from the relation
\begin{equation}
\langle z|p_z'^- \rangle = \langle z|- p_z'^+\rangle^*.
\end{equation}
Therefore, using Eqs.~(\ref{eq6}) and (\ref{eq61}), we have
\begin{equation}
\langle z| p_z'^-\rangle =   \frac{1}{2\pi}\left\{
\begin{array}{ll}
 a^{-} e^{\ii p'_z \, z}+b^{-} e^{-\ii p'_z \,z} , & z\to +\infty, \\
 c^{-} e^{\ii p'_z \, z}+d^{-} e^{-\ii p'_z \,z} , & z\to -\infty,
\end{array}\right. 
\label{eq9}
\end{equation}

\begin{equation}
\begin{split}
& p'_z>0 : a^-=1, \ d^- = 0, \\
& p'_z<0 : c^-=1, \ b^- = 0.
\end{split}
\label{eq91}
\end{equation}
Equation~(\ref{eq6}) together with the lower line of Eq.~(\ref{eq61}) describes the electron incident on the
surface, as in the low energy electron diffraction (LEED) experiment. Interestingly, the wave function of
Eq.~(\ref{eq9}) together with the upper line of Eq.~(\ref{eq91}) is a time-reversed LEED state. This kind of
function describes the photoelectron (PE) final state in the one-step theory of photoemission
\cite{FeibelmanEastman-74}. Note that in vacuum it contains both outgoing and incoming
beam. Thus, while in LEED and PE setups each of these kinds of the 
wave-functions enters separately, in EELS they are present together.

For $\rho_{ext}(\rv)$ we can write
\begin{equation}
\rho_{ext}(\rv)= \rho_{ext}(z) \frac{e^{\ii \qv_\| \cdot\rv_\|}}{(2\pi)^2},
\end{equation}
where 
\begin{equation}
\rho_{ext}(z)= \langle z|p_z^+ \rangle^* \langle z|p_z'^- \rangle,
\label{rhoz}
\end{equation} 
and $\qv_\| \! = \! \pv_\|-\pv_\|'$.
Then, finally, Eq.~(\ref{dsdwdW}) takes the convenient form
\begin{widetext}
\begin{equation}
\frac{1}{A} \frac{d^2 \sigma}{d \omega d \Omega} (\pv'\leftarrow \pv) = 
-\frac{ p'  }{\pi p \, q_\|^2} \, \text{Im} \int \rho_{ext}^*(z'') e^{-q_\||z''-z|} \chi(z,z',q_\|,\omega)  e^{-q_\||z'-z'''|} \rho_{ext}(z''')  
   d z d z' d z'' d z''',
\label{dsdwdW1}
\end{equation}
\end{widetext}
where $A$ is the surface normalization area.

\

\subsection{Real-space solution with the background dielectric function}
\label{rss}
\begin{figure} [h!] 
\includegraphics[width=  \columnwidth, trim= 75 60 35 45, clip=true]{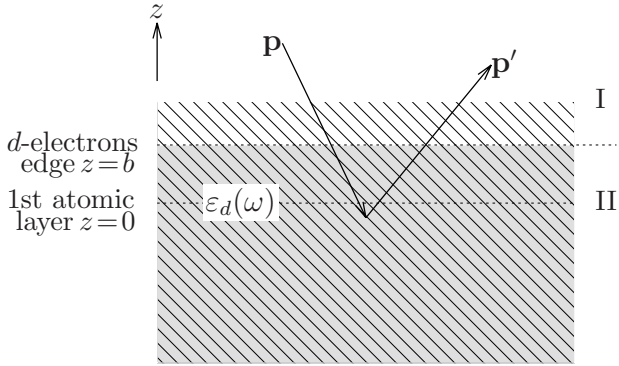}
\caption{\label{geometry} Schematics of the model used in the calculation.
}
\end{figure}

According to Eq.~(\ref{dsdwdW1}), the external potential applied to our system is
\begin{equation}
\phi_{ext}(z)= \frac{2\pi}{q_\|} \int e^{-q_\||z-z'|} \rho_{ext}(z') d z'.
\end{equation}
Therefore, Eq.~(\ref{dsdwdW1}) can be rewritten as
\begin{equation}
\frac{1}{A} \frac{d \sigma}{d \omega d \Omega} (\pv'\leftarrow \pv) \! = \! 
-\frac{ p'  }{4 \pi^3 p} {\rm Im} \! \int \! \rho_{ext}^*(z'') \phi_{ind}(z) d z,
\label{dsdwdW2}
\end{equation}
where
\begin{equation}
\phi_{ind}(z)= \frac{2\pi}{q_\|} 
\int e^{-q_\||z'-z|} \chi(z',z'',q_\|,\omega)  \phi_{ext}(z'') d z' d z''
\end{equation}
is the potential induced in the system in response to the external charge-density $\rho_{ext}(z)$. To determine $\phi_{ind}(z)$,
a simplified model of Ag surface, introduced by Liebsch \cite{Liebsch-95,Liebsch-98},  is used, in which only $s$-electrons are treated quantum-mechanically through the calculation of their response function,
while the influence of $d$-electrons is included effectively by the use
of a background  DF
$\epsilon_d(\omega)$ comprising the half-space  $z\le b$, as schematized in Fig.~\ref{geometry}.
Then, for the total scalar potential $\phi(z)=\phi_{ext}(z)+\phi_{ind}(z)$ 
we can write separately in the regions I of $z\ge b$ and  II
of $z\le b$ 
\begin{equation}
\phi(z)= \left\{
\begin{array}{ll}
 \tilde \phi(z) +A e^{-q_\| z}, & z\ge b, \\
 \dfrac{\tilde\phi(z)}{\epsilon_d}  + B e^{q_\| z}, & z\le b,
\end{array}
\right.
\label{phi}
\end{equation}
where
\begin{equation}
\tilde \phi(z) = \phi_{ext}(z)+\phi_s(z),
\label{phit}
\end{equation}
$\phi_s(z)$ is the potential of the response of $s$-electrons only,
and $A$ and $B$ are constants to be determined from 
the boundary conditions of the continuity of the tangential component of the electric field
and the normal component of the electric displacement vector, which give, respectively,
\begin{equation}
\begin{split}
&\tilde \phi(b) +A e^{-q_\| b}= \frac{\tilde\phi(b)}{\epsilon_d}  + B e^{q_\| b},\\
&-A e^{-q_\| b}=  B e^{q_\| b} \epsilon_d.
\end{split}
\label{eq18}
\end{equation}

For $\phi_s$ we can write
\begin{equation}
\begin{split}
\phi_s(z)  &=  \frac{2\pi}{q_\|} \int  e^{\ -q_\| |z  -  z'|} 
\chi_s(z', z'', q_\|, \omega) \\ &\times \left[ \phi(z'')  - \phi_s(z'') \right] 
d z' d z''  ,
\end{split}
\label{phip}
\end{equation}
where $\chi_s$ is the density-response function of $s$-electrons only.
By further rewriting Eq.~(\ref{phi}) as
\begin{equation}
\begin{split}
&\phi(z)  -  \phi_s(z)  = \\ & \left\{ \!
\begin{array}{lll}
 \phi_{ext}(z) +A e^{-q_\| z}, & z\ge b, \\
 \dfrac{\phi_{ext}(z)}{\epsilon_d} \! + 
 \! \left( \dfrac{1}{\epsilon_d} \! - \! 1 \right) \phi_s(z) \! + \!
  B e^{q_\| z} & z\le b,
\end{array}
\right.
\end{split}
\label{eq20} 
\end{equation}
substituting Eq.~(\ref{eq20}) into the right-hand side of Eq.~(\ref{phip})
and Eq.~(\ref{phit}) into Eqs.~(\ref{eq18}), we arrive at a closed system of equations for $\phi_s(z)$,
$A$, and $B$, which is numerically solved on a grid of $z$.
Then from $\phi_s(z)$ we obtain  $\tilde \phi(z)$ by Eq.~(\ref{phit}), $\phi(z)$ by Eq.~(\ref{phi}), and 
$\phi_{ind}(z)$ as $\phi(z)-\phi_{ext}(z)$. The latter is finally used in Eq.~(\ref{dsdwdW2})
to calculate the EEL spectrum.

\section{Calculations, results,  and discussion}
\label{res}

Our  calculation of the ground-state
of the $s$-electrons of the Ag (111) uses 
the 1D interpolation of the  surface and the bulk potential of Ref.~\onlinecite{Chulkov-99}.
A super-cell with the period $d=221.7$ a.u. was used,
which included 31 layers of the model $s$-subsystem of Ag, the rest occupied with vacuum.
The time-dependent density-functional theory (TDDFT) calculation of the density-response function 
$\chi_s(z,z',q_\|,\omega)$ 
is performed on the level of the random-phase approximation (RPA), i.e., setting the exchange-correlation
kernel \cite{Gross-85} $f_{xc}$ to zero. Then we apply the procedures of Secs.~\ref{rss} and \ref{flat}
to account for the response of $d$-electrons and to finally obtain the EEL spectra.
The edge of the $d$-electrons was set at $b=0.717$ a.u. above the upper atomic layer.
For $\varepsilon_d(\omega)$ we take
\begin{equation}
\varepsilon_d(\omega)=\varepsilon_{\text{Ag}}(\omega)-\left[\varepsilon_s(\omega)-1\right]=
\varepsilon_{\text{Ag}}(\omega)+\frac{\omega_{p}^2}{\omega^2},
\label{epsd}
\end{equation}
where $\varepsilon_{\text{Ag}}(\omega)$ is the experimental optical DF
of silver\cite{Palik} and $\varepsilon_s(\omega)=1-\omega_{p}^2/\omega^2$ is the Drude
DF of $s$-electrons with the plasma frequency $\omega_{p}=9$ eV.

\begin{figure} [h!] 
\includegraphics[width=  \columnwidth, trim= 60 0 10 0, clip=true]{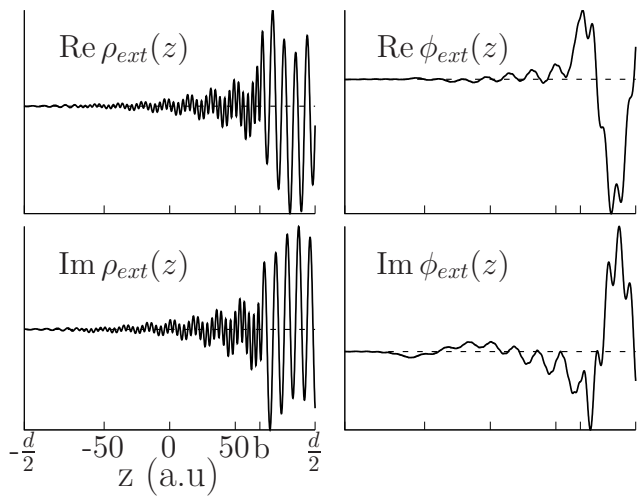}
\caption{\label{ext}
The complex-valued density of Eq.~(\ref{rhoz}) (left panels) and the corresponding
potential (right panels) as a function of the coordinate $z$ at $\omega=3.7$ eV, 
$\theta_i=80^\circ$, $\theta_s=75^\circ$, and $E_p=40$ eV. At these parameters, $q_\|=0.111$ a.u.
The edge of the $d$-electrons is at $b=67.2$ a.u. and the period of the super-cell
used in the calculation of the response of $s$-electrons is $d=221.7$ a.u.
In the calculation, the origin is chosen in the center of the super-cell.
}
\end{figure}

To construct  $\rho_{ext}(z)$  by Eq.~(\ref{rhoz}), we were solving the Schr\"{o}dinger equation 
with the asymptotic boundary conditions of Eqs.~(\ref{eq6}) and (\ref{eq9}). 
The scattering wave functions were obtained by solving the inverse band-structure problem 
as explained in Ref.~\onlinecite{Krasovskii1997} and matching the Bloch solutions in the crystal
to the linear combination of the incident and reflected wave in the vacuum. The same  crystal 
potential as for the evaluation of $\chi$ was used, with the addition of the absorbing imaginary
potential $-\ii V_{\rm i}$ as explained below.
Importantly, similar to the low-energy electron diffraction (LEED) theory~\cite{Krasovskii2002}, 
the inclusion of the optical potential (OP) into the Hamiltonian is necessary for  EELS theory as well.
This can be understood considering that, without OP, electrons having gone an arbitrarily long 
round-trip into the depth of the sample, 
would contribute to the spectrum. Since, in the first Born approximation, the probability of the bulk energy-loss is
proportional to the path-length travelled,  this would make the intensity of the bulk losses infinitely high.
The influence of the deep interior of the sample is, however, suppressed,
in LEED by all the inelastic processes,  and in EELS by the inelastic processes
beyond the first Born approximation.
In the present calculation $V_{\rm i}$ 
was taken to be spatially constant in the solid and zero in vacuum. At $E_p=40$~eV it was $V_{\rm i}=0.3$~eV
for the angle with the normal to the surface of $80^\circ$, $75^\circ$, $70^\circ$,
$V_{\rm i}=0.5$~eV for $85^\circ$, and $V_{\rm i}=0.1$ eV for $55^\circ$. 
In Fig.~\ref{ext} $\rho_{ext}(z)$ and the corresponding $\phi_{ext}(z)$ are shown for representative values of the parameters
of the EELS experiment.

\begin{figure} [h!] 
\includegraphics[width=  \columnwidth, trim= 50 0 40 0, clip=true]{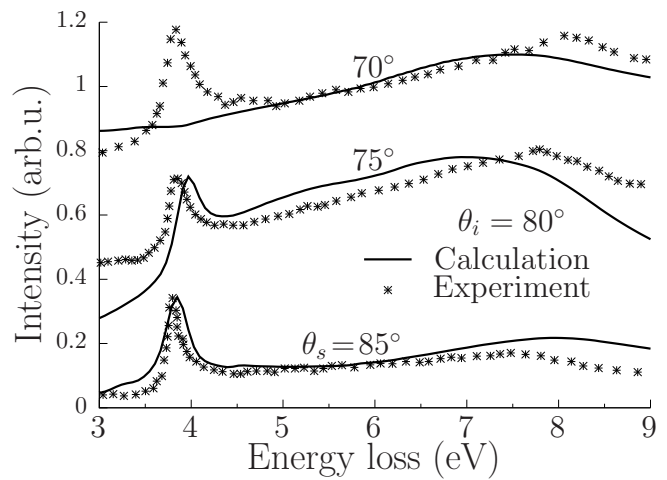}
\caption{\label{sp1}
HREEL spectra of Ag (111) surface calculated within the framework of our theoretical approach (solid lines)
compared to the experimental spectra of 10 monolayers of Ag on the Ni (111) surface, the latter
compiled from Ref.~\onlinecite{Politano-13-3} (symbols). The energy of electrons in the incident beam is $E_p=40$ eV.}
\end{figure}
\begin{figure} [h!] 
\includegraphics[width=  \columnwidth, trim= 50 0 40 0, clip=true]{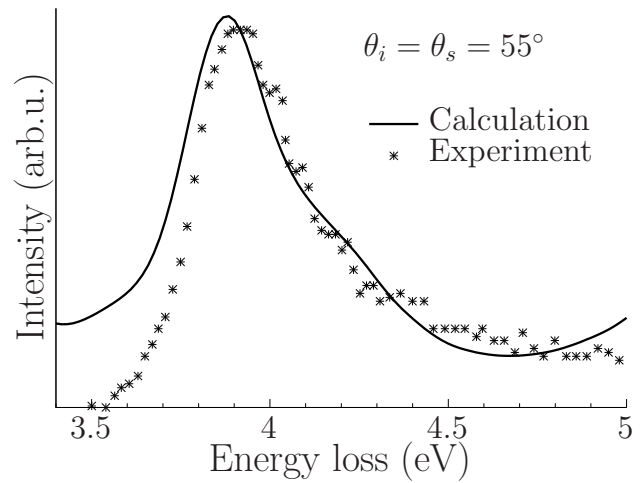}
\caption{\label{sp2}  
The same as Fig.~\ref{sp1}, but for the specular reflection $\theta_i=\theta_s=55^\circ$.
}
\end{figure}

In Figs.~\ref{sp1} and \ref{sp2} results of  calculations of the EEL reflection spectra
are presented.
They are compared to experimental HREELS of the system of 10 monolayers of Ag on (111) surface of 
Ni substrate \cite{Politano-13-3}.
In Fig.~\ref{sp1} the theoretical and experimental EEL spectra are shown for the primary energy of electrons
$E_p=40$ eV, the angle of incidence $\theta_i=80^\circ$, and three values of the angle of scattering $\theta_s$ of $85^\circ$, $75^\circ$, and $70^\circ$. In Fig.~\ref{sp2} the results for the specular geometry with $\theta_i=\theta_s=55^\circ$ and the same primary energy are presented. 
The comparison of the theory with the experiment is reasonably good. Most importantly,
the sharp bulk and surface plasmon peaks near 3.7 eV, separately present in the plots of the corresponding energy-loss functions 
(Fig.~\ref{sp0}),
are never resolved from each other in our calculations, but they form a joint broadened peak 
with a contribution from the both types of excitations. This is in 
full agreement with the HREELS experiments \cite{Rocca-95-2,Rocca-95,Moresco-97,Politano-13-3}.
A notable exception from the agreement between the theory and the experiment is 
the case of $\theta_i=80^\circ$ and $\theta_s=70^\circ$, upper spectrum in Fig.~\ref{sp1}, when, surprisingly, the lower-energy plasmon
peak, which is strong in the experimental spectrum, is absent in the theoretical one. 

\begin{figure} [h] 
\includegraphics[width=  \columnwidth, trim= 50 0 40 0, clip=true]{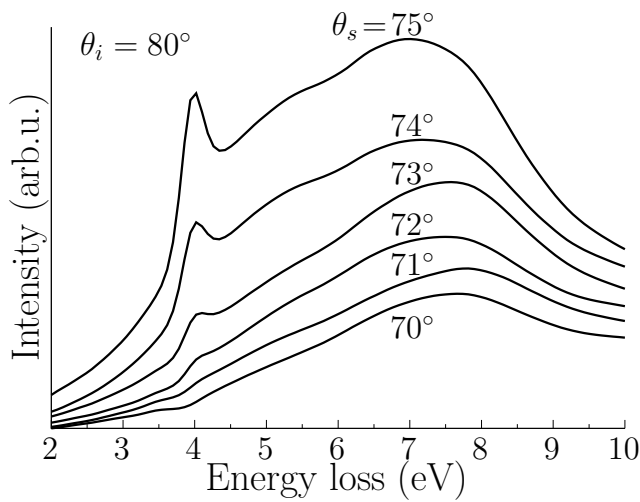}
\caption{\label{sp3} 
The calculated EEL spectra for  the angle of scattering changing from $75^\circ$ to $70^\circ$
with the step of $1^\circ$. The gradual disappearance of the peak around 3.7 eV can be observed.
}
\end{figure}

To examine the latter discrepancy closer, in Fig.~\ref{sp3}
we plot theoretical spectra for the angle of scattering gradually changing from $75^\circ$,
when the peak in question is pronounced, to $70^\circ$, when this peak disappears.
These results  show  that the strength of the peak near 3.7 eV decreases systematically when the
scattering angle  $\theta_i-\theta_s$ increases. We note that a similar effect
of the disappearance of the 3.7 eV peak with the growing momentum
can be observed in the results of the calculations of Ref.~\onlinecite{Silkin-15-3},
performed  in the dipole-scattering mode within the same model of $d$-electrons.
We analyze this tendency in detail in Appendix \ref{prop} to the conclusion that, although,
the background DF model is applicable in the higher energy range around 8 eV
to account for the bulk, surface, and the multipole plasmons in Ag\cite{Liebsch-95,Liebsch-98},
it fails for the lower-energy plasmons 
at larger values of the wave-vector. 
Obviously, the future theory, which will include $d$-electrons from the first principles,
will be free from this deficiency. 

\begin{figure} [h] 
\includegraphics[width=  \columnwidth, trim= 50 0 40 0, clip=true]{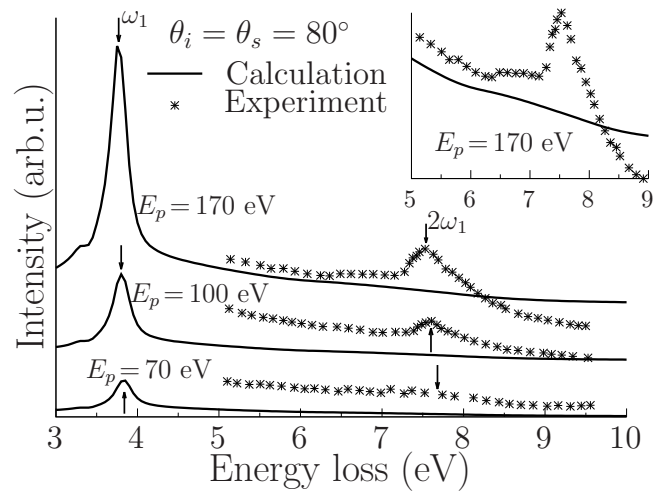}
\caption{\label{8080} 
Primary energy dependence of HREELS of Ag (111).
Peaks on the experimental spectra are attributed
to the double-plasmon excitation and are, accordingly, absent at the theoretical spectra.
The inset shows the same spectra for $E_p=170$ eV in a narrower energy range.
The experimental data compiled from Ref.~\onlinecite{Politano-13-3} are used.
}
\end{figure}

Since our calculations based on Eq.~(\ref{dsdwdW}) are linear with respect to the interaction between the probe electron and the electronic subsystem of the target (distorted-wave\cite{Taylor} with the first Born approximation for the inelastic processes),
the multiple energy losses, e.g., multiple plasmon excitations, are beyond the capacity of this approach.
Nonetheless, especially at higher primary energies, multiple losses can be expected in the experimental spectra.
In Fig.~\ref{8080} we plot the theoretical spectra together with the corresponding 
experimental ones for three different primary energies of 170, 100, and 70 eV at the specular geometry of $\theta_i=\theta_s=80^\circ$,
where we now focus on the higher energy range.
While the theoretical lines are rather smooth in this range, the experimental spectra 
at $E_p=$ 170 and 100 eV have prominent peaks around 7.6 eV.
Considering that (i) the positions of these peaks are very close to the twice the energy of the strong 
single-plasmon peaks around 3.7 eV, (ii) their intensities  change with $E_p$ consistently with those of the corresponding
single-plasmon peaks, and (iii) these peaks are present in the experiment while
absent in the linear-response based calculations,
we are led to the conclusion that these peaks are due to the double-plasmon excitations.

To reproduce the multiple plasmons theoretically, a theory of EELS beyond the first Born approximation 
is required.
We note that the formal theory of inelastic scattering of a quantum-mechanical particle
to the {\em second} Born approximation, expressed in terms of the {\em quadratic} density-response function
of a target, was constructed in Ref.~\onlinecite{Nazarov-02}. 
The practical implementations of this theory have not, however, been yet developed.

\section{Conclusions}
\label{concl}
We have revisited the problem of the energy losses by electrons
in reflection electron energy-loss spectroscopy with the focus
on the role of the probing electrons' kinematics. The inadequacy of the description of EELS
in terms of the energy-loss functions has been emphasized  for the materials 
where the bulk and the surface features overlap or are close in energy in the excitation spectra.
We have implemented the theory including the effect of the detailed quantum-mechanical motion 
of the probe electrons during their energy losses, using Ag surface as a representative example of a system
where the kinematic  aspect of the problem is particularly important.

Since our primary interest lies in the role of the kinematic effects, 
for a sophisticated problem of the $d$-electrons' response we have used a simplified model of
the background dielectric function, which has immensely simplified the numerical implementation.
As a side effect, although we have found a reasonably good overall agreement with HREELS, 
at  strongly off-specular geometries of the  experiment, the theory and the measurements disagree.
We have tracked this discrepancy down to the failure  of the 
substitution of the $d$-electrons with the background dielectric function 
to describe the dispersion of the main loss feature of 3.7 eV 
in Ag at larger values of the momentum. Thus, the limits of the applicability of that, otherwise very useful model, have been set.
This difficulty is anticipated to be overcome in the future theory, with all the ingredients
included within the {\em ab initio} approach.

Another deviation of the theory from experiment we have found can be qualified
as an evidence of the consistency of the former rather than its deficiency. 
Namely, the experiment shows
peaks at EEL spectra that, by all the evidence, can be attributed to the double-plasmon excitations.
The linear-response theory, which our approach is based on, fundamentally cannot account 
for such losses, and we do not, accordingly, obtain the double-excitation peaks in the calculations.
Future implementations of the theory of the quadratic and higher-order response 
will be able to account for these processes.

\acknowledgments
VUN acknowledges support from the Ministry of Science and Technology, Taiwan, Grant  \mbox{104-2112-M-001-007}.
This work was supported by the Spanish Ministry of Economy and
Competitiveness MINECO (Project No. FIS2013-48286-C2-1-P). 

\appendix

\section{Properties of the model of the background DF}
\label{prop}

In this Appendix we scrutinize  the  model of the background DF
used in this paper to account for the  $d$-electrons in the dielectric response of silver\cite{Liebsch-95,Liebsch-98}.
For the sake of maximal clarity, we do this analytically by considering the bulk response.
In this case the same model is determined by the DF 
\begin{equation}
\begin{split}
\varepsilon(q,\omega) &=\varepsilon_{\text{Ag}}(\omega)-\left[1-\frac{\omega_p^2}{\omega^2}\right]
+  \varepsilon_{Ls}(q,\omega) \\
&= \varepsilon_d(\omega)+\left[ \varepsilon_{Ls}(q,\omega)-1 \right],
\end{split}
\label{epsb}
\end{equation}         
where $\varepsilon_{\text{Ag}}(\omega)$ is the optical experimental DF of Ag,
$\varepsilon_d(\omega)$ is given by Eq.~(\ref{epsd}),
and $\varepsilon_{Ls}(q,\omega)$ is the Lindhard's DF
of the homogeneous electron gas\cite{Lindhard-54}, taken at the density of the $s$-electrons of Ag.
         
\begin{figure} [h] 
\includegraphics[width=  \columnwidth, trim= 35 0 40 0, clip=true]{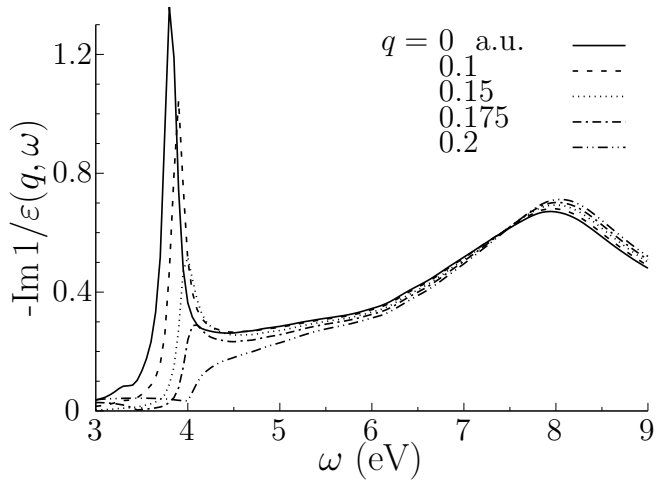}
\caption{\label{bulk} 
The bulk energy-loss function using the DF of Eq.~(\ref{epsb})
plotted at several values of the wave-vector $q$. The plasmon peak near 3.7 eV clearly
weakens and finally disappears with the increase of $q$.
}
\end{figure}        

\begin{figure} [h] 
\includegraphics[width=  \columnwidth, trim= 30 0 40 0, clip=true]{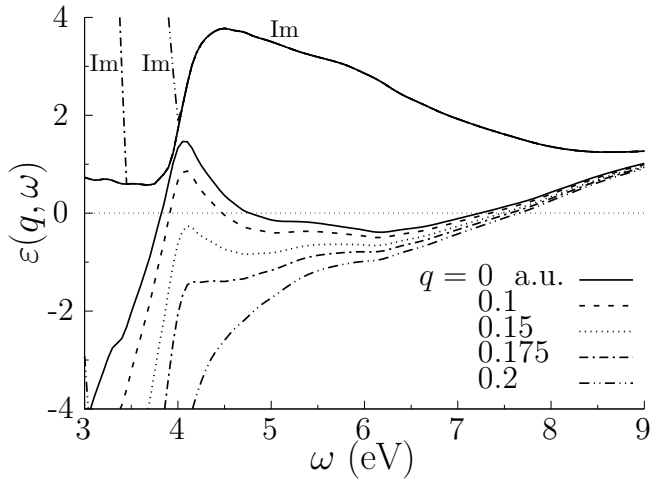}
\caption{\label{bulke} 
The dielectric function of Eq.~(\ref{epsb}). Lines without labels are the real parts of the DF
at the corresponding values of $q$.
Lines with the label ``Im'' are imaginary parts, which, for all values of $q$,
largely coincide with $-\text{Im}\,\varepsilon(q=0,\omega)$,
since the Lindhard's DF is real in the corresponding energy ranges.
}
\end{figure}

In Fig.~\ref{bulk} we plot the energy-loss function with the use of the DF of Eq.~(\ref{epsb})
for several values of the wave-vector $q$. The plasmon peak near 3.7 eV weakens with the increase 
of $q$, until it disappears at $q\approx$ 0.2 a.u. This is consistent with the behavior 
of the  DF itself plotted in Fig.~\ref{bulke}. Indeed, 
for $q=$ 0 and 0.1 a.u.,  Re $\varepsilon(q,\omega)$ crosses zero in the corresponding energy range,
thereby producing the plasmon peak in the loss-function.
This is not the case any more for $q=$ 0.15, 0.175, and 0.2 a.u.,
although, for the former two wave-vectors, the  peaks in question still persist in the loss-function
due to the  Re\ $\varepsilon$ {\em approaching}  the zero axis (cf., Ref.~\onlinecite{Nazarov-15}). Lastly, at $q=$ 0.2 a.u., 
Re\ $\varepsilon$ is very far from zero in this $\omega$-range, and no peak in the loss-function can
be discerned any more. 

The above results are consistent with those of Sec.~\ref{res} of this paper. Indeed, 
the main contribution to the $z$-component of the wave-vector of the external perturbation can be estimated 
from Eqs.~(\ref{rhoz}), (\ref{eq6}), and (\ref{eq9}) as $p_z+p'_z$ (note that $p_z<0$ and $p'_z>0$).
When $E_p=$ 40 eV and $\theta_i=$ 80$^\circ$, this is equal to 0.123 and 0.259 a.u.,  for $\theta_s=$ 75$^\circ$
and 70$^\circ$, respectively. The corresponding values of $q_\|$ are 0.117 and 0.160 a.u., respectively.
Then $q=[(p_z+p'_z)^2+q_\|^2]^{1/2}$ are 0.170 and 0.304 a.u., respectively, explaining the presence
of the plasmon near 3.7 eV in the theoretical spectra in the former and its absence in the latter case.
For the specular geometry in Fig.~\ref{sp2}, the corresponding value is $q=$ 0.09 a.u.,
consistent with the strong theoretical lower-energy plasmon peak in this figure.
The same argument holds for the spectra in Fig.~\ref{8080}.

Experimentally, however, this prediction of the model is not supported, 
as can be seen in Fig.~\ref{sp1}, upper experimental spectrum: the plasmon near 3.7 eV persists in the measurements on
Ag at larger wave-vectors.
We, therefore, can conclude that, at larger wave-vectors, the 
response of $d$-electrons  cannot be realistically substituted 
with a wave-vector independent dielectric function, when it concerns
the lower-energy plasmon in Ag, while this model is quite successful in the higher
energy range, where the spectra are dominated by the response of $s$-electrons outside
the $d$-electrons background\cite{Liebsch-95,Liebsch-98}.
                                       
%

\end{document}